\documentclass[twocolumn,showpacs,preprintnumbers,amsmath,amssymb]{revtex4}
\usepackage{graphicx}
\usepackage{dcolumn}
\usepackage{bm}

\newcommand{\bef}{\begin{figure}}
\newcommand{\eef}{\end{figure}}

\newcommand{\be}{\begin{equation}}
\newcommand{\ee}{\end{equation}}
\newcommand{\bea}{\begin{eqnarray}}
\newcommand{\eea}{\end{eqnarray}}
\begin{document}

\title{Systematic study of symmetric cumulants at $\sqrt{s_{NN}}$ =
  200 GeV in Au+Au collision using transport approach}

\author{Md. Nasim}
\affiliation{University of California Los Angeles, CA 90095, USA}

\date{\today}
\begin{abstract}
Measurement of 4-particle symmetric cumulants has
been considered to be a good tool to study the correlations between
amplitudes of different orders of
anisotropic flow  harmonics in the heavy-ion collision. These new
observables not yet been measured at RHIC. Using   A Multi-Phase
Transport model, a set of predictions for the centrality dependence of the
normalized  4-particle symmetric cumulants
in Au+Au collisions at $\sqrt{s_{NN}}$ = 200 GeV has been given.
In addition, the  effects of shear viscosity and hadronic rescattering on the magnitude
of symmetric cumulants are discussed using AMPT model at
$\sqrt{s_{NN}}$ = 200 GeV. It is shown that $sc(2,3)$ is found to be more
sensitive to  hadronic rescattering, whereas $sc(2,4)$ is more
sensitive to the shear viscosity. Rapidity dependence of  symmetric
cumulants is also shown. A relation between symmetric cumulant and event
plane correlation is investigated using AMPT model.
\end{abstract}
\maketitle

\section{Introduction}
Elliptic flow ($v_{2}$) measured in heavy-ion collisions is believed to 
arise because of the pressure gradient developed when two nuclei collides
at non-zero impact parameter followed by subsequent interactions among
the constituents~\cite{flow1,flow1a}. If the nuclear overlap region were
smooth, only even order flow harmonics ($v_{2},v_{4},v_{6}$ etc.) would be present in the final particle
distributions. However, the nucleus is made up from a finite number of
nucleons whose positions can fluctuate considerably event-by-event
leading to fluctuations in the collision geometry~\cite{vnodd}. These
fluctuations could result in the production of odd order
eccentricities in the initial geometry leading to formation of odd
flow harmonics in the final particle distribution.  The magnitude of $v_{n}$ has been
shown to be sensitive to the initial state and the equation of state of the system formed in the 
collisions~\cite{flow2}. Event-by-event measurement of anisotropic flow is crucial to
understand the initial conditions in heavy-ion collision.\\
Correlations between different order flow harmonics are predicted to be sensitive to the transport properties of the produced medium in heavy-ion collisions.
Recently, a new tool, namely 4-particle symmetric cumulants~\cite{scmn_1},
is emerging with a promise to throw additional light on the
initial-state phenomena and the transport properties of the produced medium in heavy-ion collisions~\cite{scmn_2,scmn_3,scmn_4} .\\
The azimuthal distribution ($\phi$) of particles in a given event is written as
\begin{equation}
P(\phi)=\frac{1}{2\pi}\sum^{n=+\infty}_{n=-\infty}{V_{n}e^{-in\phi}},
\end{equation}
where $V_{n}=v_{n}e^{in\psi_{n}}$ is the $n^{th}$ harmonic anisotropic
flow coefficient with respect to event plane angle $\psi_{n}$. The
4-particle symmetric cumulants $SC(n,m)$ with $n\neq m$~\cite{scmn_1} can be defined as 
\begin{equation}
SC(n,m)\equiv \langle v_{n}^{2}v_{m}^{2} \rangle - \langle
v_{n}^{2}\rangle  \langle v_{m}^{2} \rangle. 
\end{equation}
Normalized symmetric cumulants $sc(n,m)$ is defined as
\begin{equation}
sc(n,m)\equiv \frac{\langle v_{n}^{2}v_{m}^{2} \rangle - \langle
  v_{n}^{2}\rangle  \langle v_{m}^{2} \rangle}{\langle
  v_{n}^{2}\rangle  \langle v_{m}^{2} \rangle}. 
\label{scmn_norm}
\end{equation}
Magnitude of $sc(n,m)$ gives correlation strength between  $\langle
v_{n}^{2}\rangle $ and $ \langle v_{m}^{2}\rangle $. \\
Recently, the ALICE collaboration has measured the $sc(2,3)$ and
$sc(2,4)$ as a function of collision centrality~\cite{scmn_2}. This
measurement has attracted an increased attention of many physicists, since a simultaneous description of
$v_{n}$ and $sc(n,m)$ cannot be captured using a single model with a constant initial condition and transport coefficient.
Such measurement not yet been done at RHIC, leaving an opportunity to
make predictions. In this paper, I made prediction of $sc(n,m)$ for upcoming
measurements at RHIC as well as I have systematically studied the magnitude of  $sc(n,m)$ under
various condition using a transport model. This study will give a base line
to understand the experimental data as well as a test of AMPT model
which is very successful in describing magnitude of flow harmonics at
RHIC energies.\\

This paper is organized in the following way.
Section 2 describes details of the model used.
 In Section 3, transverse momentum spectra and magnitude of
 anisotropic flow harmonics of charged particle from AMPT model is
 presented. Comparisons with data and model are also shown.  
 In Section 4, the effect of shear viscosity and hadronic re-scattering on the magnitude
of $sc(n,m)$ are discussed. Rapidity dependence of $sc(n,m)$ is also
shown in Section 4. Sections 5 describes relation between $sc(n,m)$
and event plane correlation. Finally, I summarize in Section 5.

\section{AMPT Model}
 The AMPT model is a hybrid transport model~\cite{{ampt}}. It uses the same initial
 conditions as in HIJING~\cite{hijing}. The AMPT model can be studied in two
 configurations, in the AMPT default version (AMPT-Def)  in which the minijet
 partons are made to undergo scattering before they are allowed to
 fragment into hadrons~\cite{lund}, and in the AMPT string melting
 scenario ( AMPT-SM)  where
 additional scattering occurs among the quarks and hadronization
 occurs through the mechanism of parton coalescence. The string melting  version
 of the AMPT model is based on the idea that for energy densities
 beyond a critical value of $\sim$1$GeV/fm^{3}$, it is difficult to
 visualize the coexistence of strings (or hadrons) and partons. Hence,
 the need to melt the strings to partons. The scattering of the quarks
 is based on parton cascade~\cite{ZPC}.
 In AMPT model, the value of parton parton scattering cross-section, $\sigma_{pp}$, is calculated by 
\be
\sigma_{pp} \approx \frac{9 \pi \alpha_s ^2}{2 \mu^2},
\label{muaplha}
\ee
where $\alpha_s$ is QCD coupling constant and $\mu$ is screening mass. 
In this study, approximately 5 M Au+Au  events are generated for each
configuration.

\section{TRANSVERSE MOMENTUM SPECTRA and ANISOTROPIC FLOWS from AMPT model}
\begin{figure}
\begin{center}
\includegraphics[scale=0.4]{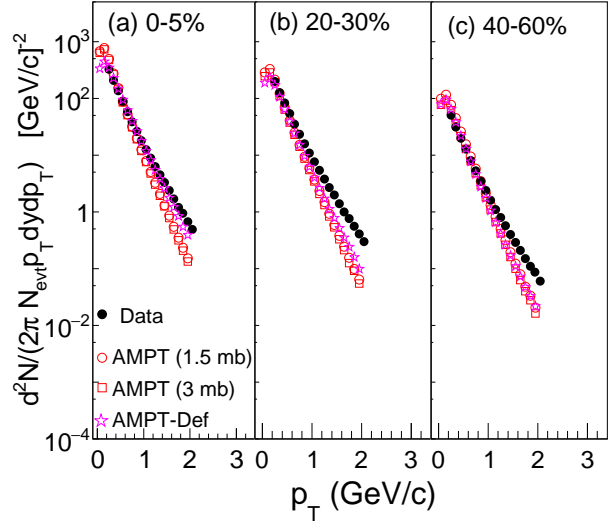}
\caption{(Color online)  Transverse momentum spectra of charged hadron
  in Au+Au collision at $\sqrt{s_{NN}}$ =200 GeV for (a) 0-5$\%$ (most
  central) (b) 20-30$\%$ (mid-central) and (c) 40-60$\%$ (peripheral)
  centrality. Black solid circles
  are data~\cite{spectra_star} and open  markers are AMPT model calculation.}
\label{spectra_ampt}
\end{center}
\end{figure}

Before I make a prediction using AMPT model, I need to fix set of
input parameters by fitting experimental data on transverse momentum spectra and anisotropic
flow harmonics.
Figure~\ref{spectra_ampt} shows the transverse momentum ($p_{T}$)
spectra of mid-pseudorapidity ( $|\eta|$ $<$ 0.5) charged particles in different
centrality bins. Black solid circles are data from STAR
experiment~\cite{spectra_star} and open  markers are AMPT model
calculations with parton-parton interaction cross-section 1.5 mb and
3mb and default configuration. It is seen that AMPT model describe
reasonably the experimental data at low $p_{T}$ for all centralities.  AMPT fails to
describe data at high $p_{T}$ ($>$ 1.5 GeV/c). This is due to the
small current quark masses used in the AMPT model 
so that partons are less affected by the radial flow effect.

\begin{figure*}
\begin{center}
\includegraphics[scale=0.85]{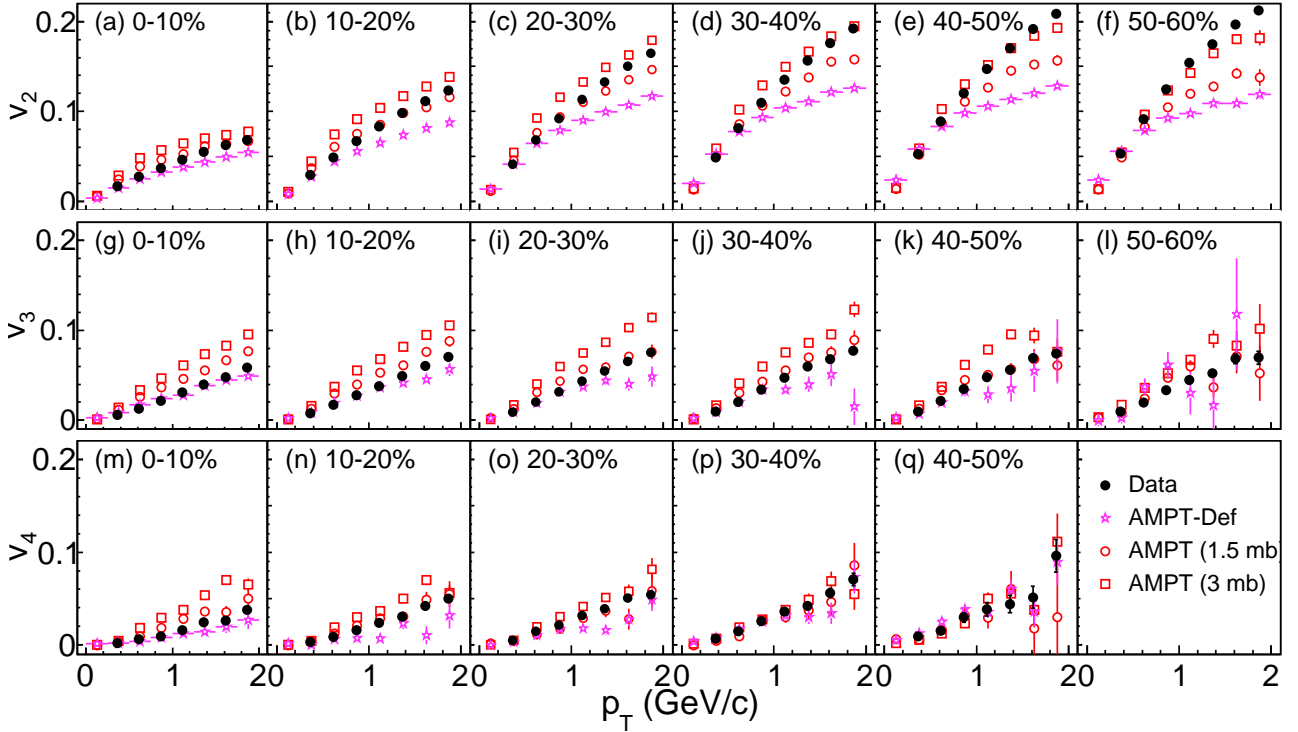}
\caption{(Color online)  Transverse momentum dependence of $v_{2}$,
  $v_{3}$, and $v_{4}$
  in Au+Au collision at $\sqrt{s_{NN}}$ =200 GeV for  0-10$\%$,
   10-20$\%$,  20-30$\%$, 30-40$\%$, 40-50$\%$, and 
  50-60$\%$ centrality. Black solid circles
  are data~\cite{vn_phenix} and open markers are AMPT model calculation.}
\label{v2_ampt}
\end{center}
\end{figure*}

Now I show  the transverse momentum  dependence of  charged
particle $v_{n}(\psi_{n})$ at mid-pseudorapidity ($|\eta|$ $<$ 0.35)  in Au+Au collisions at $\sqrt{s_{NN}}$
= 200 GeV obtained from the AMPT model and its comparison with data,
measured by PHENIX experiment~\cite{vn_phenix}. Figure~\ref{v2_ampt}
shows comparison between data and AMPT
model calculation for $v_{2}(\psi_{2})$, $v_{3}(\psi_{3})$ and
$v_{4}(\psi_{4})$, respectively. For $v_{n}$ measurement in AMPT model,
the azimuthal angle ($\phi$) of each particle is correlated with event
plane $\psi_{n}^{avg}$. Where  $\psi_{n}^{avg}$ is the average of
$\psi_{n}^{pos}$ and  $\psi_{n}^{neg}$, calculated using charged
particles within  1.0 $<$ $\eta$ $<$ 2.8 and -2.8 $<$ $\eta$ $<$ -1.0,
respectively (the same method as used in experimental data
analysis~\cite{vn_phenix}). Event plane resolutions in AMPT are
comparable with the PHENIX data within the limit of 10$\%$. It is seen
from  Fig.~\ref{v2_ampt} that
$v_{n}(\psi_{n})$ measured by PHENIX can be described by using
parton-parton cross-section between 1.5 to 3 mb. However, the
agreement between data and model is not good for most central
collisions. The AMPT model with default configuration predicts smaller $v_{n}$
compared to  AMPT with  $\sigma_{pp}$ = 1.5 and 3 mb.
\section{$sc(n,m)$ from AMPT model}
The magnitude of $ v_{n}^{2}v_{m}^{2} $ and $v_{n}^{2} $ in the numerator of
Eq.~\ref{scmn_norm} is calculated using
multi-particle cumulant method~\cite{scmn_1} as shown below.\\
\begin{small}
\begin{eqnarray}
&& v_{n}^{2}v_{m}^{2} =\frac{1}{\binom{M}{4}4!}\,\sum_{\begin{subarray}{c}i,j,k,l=1\\ (i\neq j\neq k\neq l)\end{subarray}}^{M} e^{i(m\varphi_i+n\varphi_j-m\varphi_k-n\varphi_l)}  \nonumber\\
&&=\frac{1}{\binom{M}{4}4!}\big[\left|Q_{m}\right|^2\left|Q_{n}\right|^2\!-\!
2\mathfrak{Re}\left[Q_{m+n}Q_{m}^*Q_{n}^*\right]\!  \nonumber\\ &&  -\! 
2\mathfrak{Re}\left[Q_{m}Q_{m-n}^*Q_{n}^*\right]{}\! +\!\left|Q_{m+n}\right|^2\!+\!\left|Q_{m-n}\right|^2\!
                                                                   \nonumber\\
&& -\!(M\!-\!4)(\left|Q_{m}\right|^2\!+\!\left|Q_{n}\right|^2)
+\!M(M\!-\!6)\big]\,,
\label{eq:4p}
\end{eqnarray}
and
\begin{equation}
 v_{n}^{2} =
\frac{1}{\binom{M}{2}2!}\,\sum_{\begin{subarray}{c}i,j=1\\ (i\neq j)\end{subarray}}^{M} e^{in(\varphi_i-\varphi_j)}
=\frac{1}{\binom{M}{2}2!}
\big[\left|Q_{n}\right|^2\!-\!M\big]\,.
\label{eq:2p}
\end{equation}
\end{small}
Where $M$ is the multiplicity of an event and $Q_{n}$ is flow vector for
$n^{th}$ harmonic, $Q_n\equiv\sum_{k=1}^Me^{in\varphi_k}$. 
The weights of $M(M-1)$ and $M(M - 1)(M - 2)(M - 3)$ are used to get the event-averaged 2-particle and 4-particle correlations.
The magnitude of $\langle v_{n}^{2}\rangle$ in the denominator of Eq.~\ref{scmn_norm}
is obtained with 2-particle correlations and using a pseudorapidity
gap of $|\Delta\eta| > 1.0$ to suppress biases from few-particle
non-flow correlations. Charged hadrons ($\pi$, $K$ and $p$) within 0.2
$<$ $p_{T}$ $<$ 2.0
GeV/c  and  $|\eta|$ $<$ 1.0  are used in this analysis.

\subsection{Hadronic vs. Partonic medium}
Measurement of $sc(n,m)$ has been done using both AMPT-SM (partonic
medium) and AMPT-Def (hadronic medium) models keeping all other input
parameters to be the same. The analysis was performed in 1$\%$ centrality bins,
which are then recombined into 10$\%$ bins for reducing statistical
uncertainty as suggested in ref~\cite{matt}.
Anti-correlation between $v_{2}$
and $v_{3}$ (Fig.~\ref{fig_scmn_sm_def}a) and positive correlation
between $v_{2}$ and $v_{4}$(Fig.~\ref{fig_scmn_sm_def}b) is observed for both
AMPT-SM and AMPT-Def model. However  (anti-) correlation strength
between $v_{2}$ and  ($v_{3}$) $v_{4}$ is stronger in the case of AMPT-def
model which is a pure hadronic model. If there is any measurement of
$sc(n,m)$ at BES energy at RHIC in future, this study could be
useful to understand data.
\bef
\begin{center}
\includegraphics[scale=0.35]{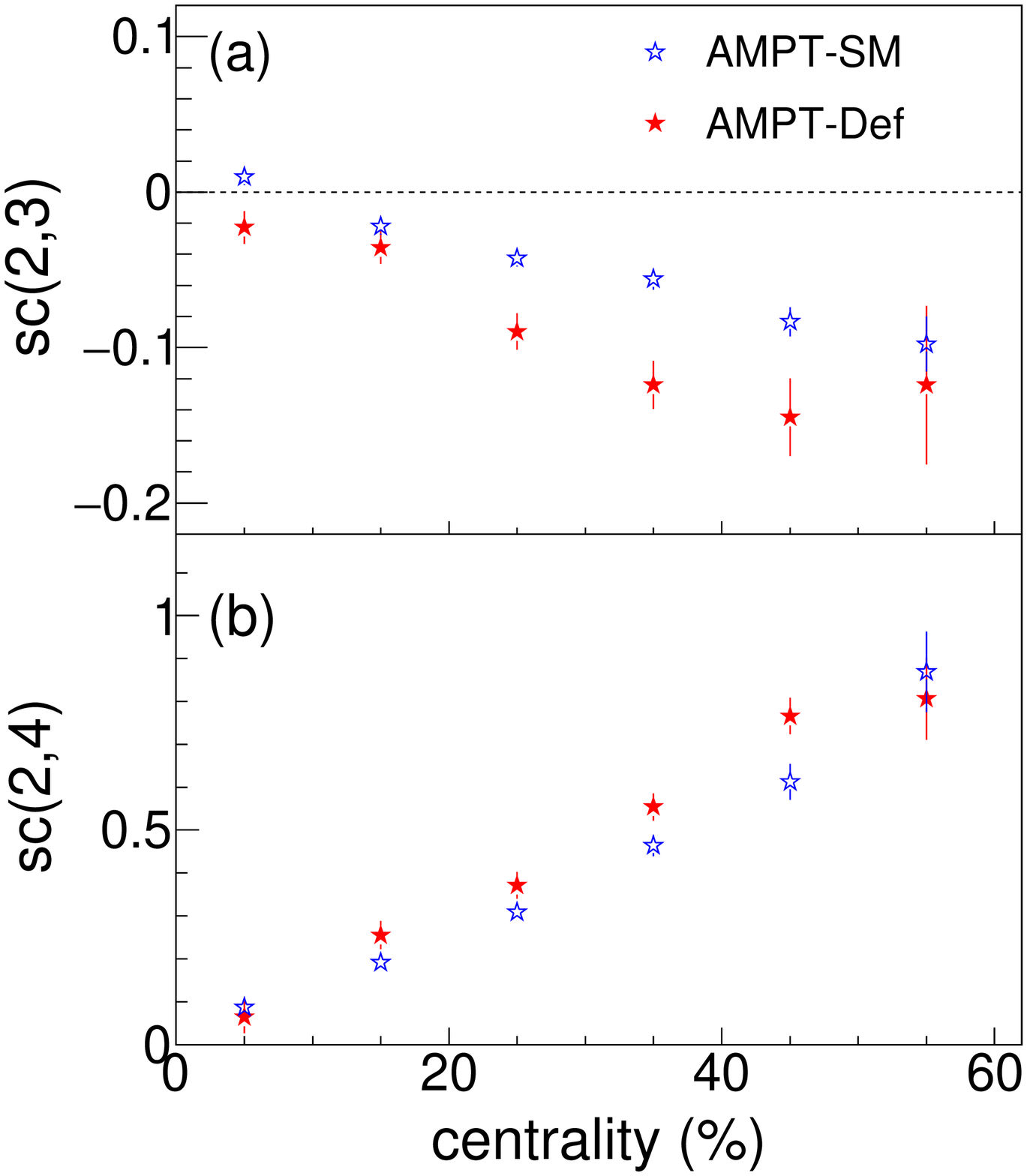}
\caption{(Color online)  Centrality dependence of (a) $sc(2,3)$ 
and (b) $sc(2,4)$  in Au+Au collision at $\sqrt{s_{NN}}$ =200 GeV from
AMPT-SM (blue) with $\sigma_{pp}$= 3 mb and AMPT-Def (red) model.}
\label{fig_scmn_sm_def}
\end{center}
\eef
\subsection{Effect of shear viscosity}
Transport coefficients play a major role in probing the properties of
the medium created in high energy heavy ion
collisions~\cite{trans_1,trans_2,trans_3,trans_4,trans_5,trans_6,trans_7,trans_8}. It
has also been predicted that magnitude of $sc(n,m)$ could be sensitive
to the transport properties of the produced medium. In this paper, the
magnitude of  $sc(n,m)$ for different values of  shear viscosity to
entropy density ratio ($\eta_s/s$ ) using AMPT-SM model at $\sqrt{s_{NN}}$ = 200 GeV is studied. For a system of massless quarks and gluons at temperature  $T$ ($T = 378$ MeV at RHIC energy in AMPT~\cite{lhc_chgv2}), the  $\eta_s/s$  is given by~\cite{lhc_chgv2}
\be
\frac{\eta_s}{s} \approx \frac{3\pi}{40 \alpha_s ^2} \frac{1}{\left( 9+ \frac{\mu^2}{T^2} \right) \ln \left( \frac{18 + \mu^2/T^2}{\mu^2/T^2} \right) - 18} 
\label{eq_visco}
\ee
 Three different value of $\eta_s/s$ e.g. 0.08,  0.18 and 0.35 keeping
 $\alpha_s =0.47$ are used in this study.
 It was shown that AMPT model with
$\eta_s/s$ between 0.18 ($\sigma_{pp}$ = 3 mb)  and 0.35
($\sigma_{pp}$ = 1.5 mb) explains magnitude of $v_{n}$.  
Fig.~\ref{fig_scmn_etabys} shows centrality dependence of $sc(2,3)$
and $sc(2,4)$ in Au+Au collision at $\sqrt{s_{NN}}$ =200 GeV from
AMPT-SM model. Black, red and blue marker corresponds to medium with
$\eta_s/s$=0.35, 0.18  and 0.08 respectively. The magnitude of $sc(2,4)$ increases with
increase in shear viscosity, however anti-correlation
between $v_{2}$ and $v_{3}$ decreases slightly with  shear viscosity.
Fig.~\ref{fig_scmn_etabys} shows that the magnitude of $sc(2,4)$  is more sensitive
to $\eta_s/s$ compared to $sc(2,3)$. One can  also finds that change in
$sc(2,4)$ above $\eta_s/s$=0.18 is negligible.
\bef
\includegraphics[scale=0.35]{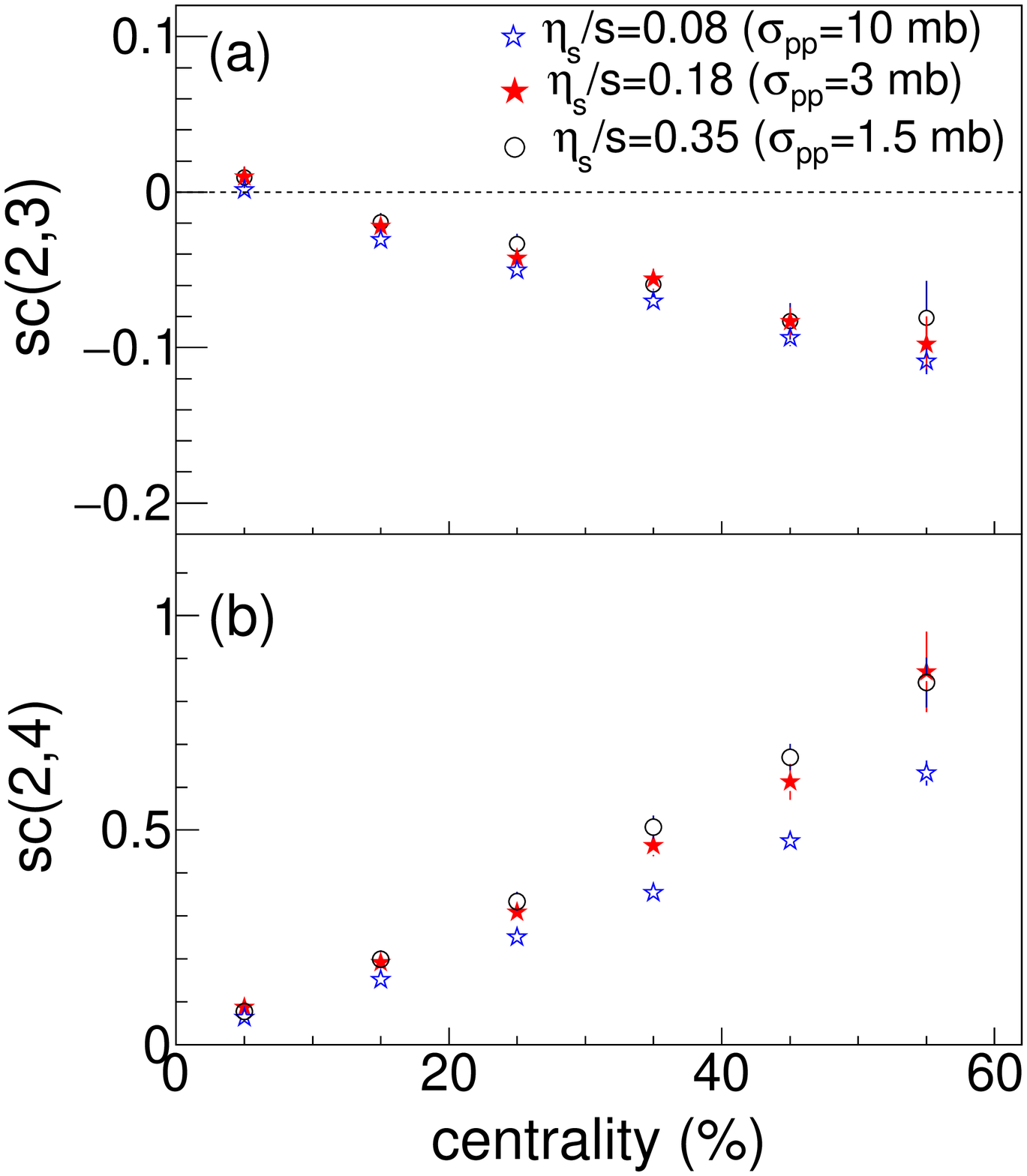}
\caption{(Color online) Centrality dependence of (a) $sc(2,3)$ 
and (b) $sc(2,4)$  in Au+Au collision at $\sqrt{s_{NN}}$ =200 GeV from
AMPT-SM model for $\eta_s/s$ = 0.35 (black), 0.18 (red)  and 0.08 (blue).}
\label{fig_scmn_etabys}
\eef
\bef
\includegraphics[scale=0.35]{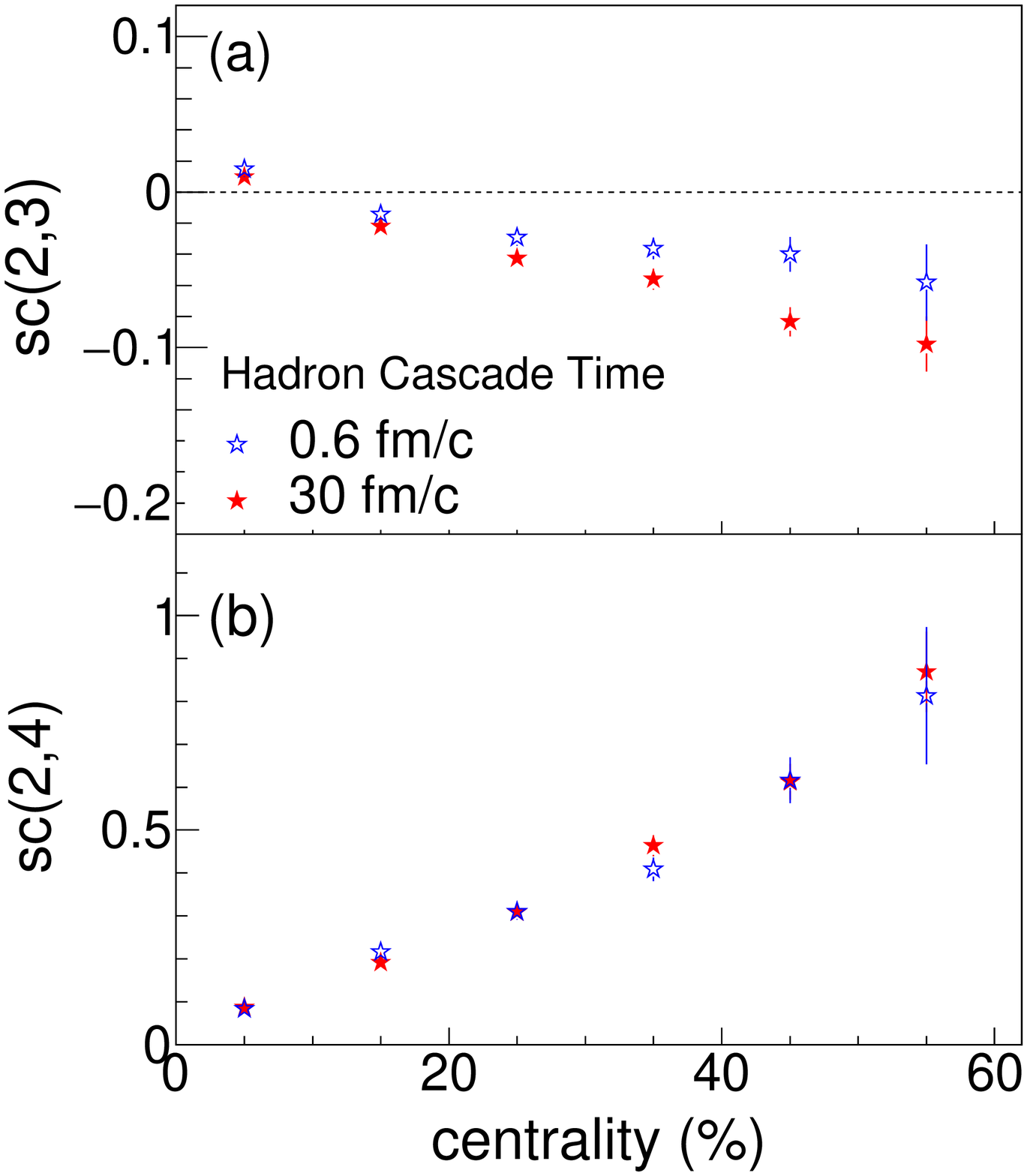}
\caption{(Color online) Centrality dependence of (a) $sc(2,3)$ 
and (b) $sc(2,4)$  in Au+Au collision at $\sqrt{s_{NN}}$ =200 GeV from
AMPT-SM model ($\sigma_{pp}$ = 3 mb) for hadron cascade time = 0.6 fm/c (blue)  and 30 fm/c (red).}
\label{fig_scmn_ntmax}
\eef
\subsection{Effect of hadronic re-scattering}
The AMPT model with string melting leads to hadron formation using a
quark coalescence model. The subsequent hadronic matter interaction is
described by a hadronic cascade, which is based on a relativistic
transport (ART) model~\cite{ART}.  The termination time of the hadronic cascade is varied in this paper from 0.6 to 30 $fm/c$ to study the effect of the hadronic rescattering on $sc(n,m)$. Higher value of hadronic cascade time reflects larger hadronic rescattering. 
Fig.~\ref{fig_scmn_ntmax} shows centrality dependence of $sc(2,3)$
and $sc(2,4)$ in Au+Au collision at $\sqrt{s_{NN}}$ =200 GeV from
AMPT-SM model with hadron cascade time = 0.6 fm/c (blue)  and 30 fm/c
(red). The magnitude of $sc(2,3)$ is found to be sensitive to hadron
cascade time or hadronic rescattering. With increases in hadronic
resecatering, the magnitude of  $sc(2,3)$ decreases indicating more
anti-correlation between $v_{2}$ and $v_{3}$. On the other hand,
change in  $sc(2,4)$ due to change in hadron cascade time is
negligible. Hence the correlation between $v_{2}$ and $v_{4}$ is not
sensitive to the hadronic rescattering.
\subsection{Rapidity dependence}
In this section, rapidity dependence of 
$sc(n,m)$ is discussed. The different experiments have different detector setup
with different rapidity coverage. Therefore, rapidity dependence of
$sc(n,m)$ is studied to give a base line for future experimental
measurement. Fig.~\ref{fig_scmn_rap} shows centrality dependence of  $sc(2,3)$ 
and $sc(2,4)$  in Au+Au collision at $\sqrt{s_{NN}}$ =200 GeV from
AMPT-SM model with pseudo-rapidity coverage $|\eta|<1.0$ (red) ,  $1 <|\eta|<3$ (black) and
$3<|\eta|<5$ (blue). A change in magnitude of
$sc(2,3)$  and $sc(2,4)$ for different pseudo-rapidity
region is observed. Anti-correlation between $v_{2}$ and $v_{3}$ is stronger at
far forward and backward pseudo-rapidity in comparison to
mid-pseudo-rapidity. A strong positive correlation between $v_{2}$
and $v_{4}$ is observed at mid-pseudo-rapidity in comparison to far forward and backward pseudo-rapidity.
\bef
\includegraphics[scale=0.35]{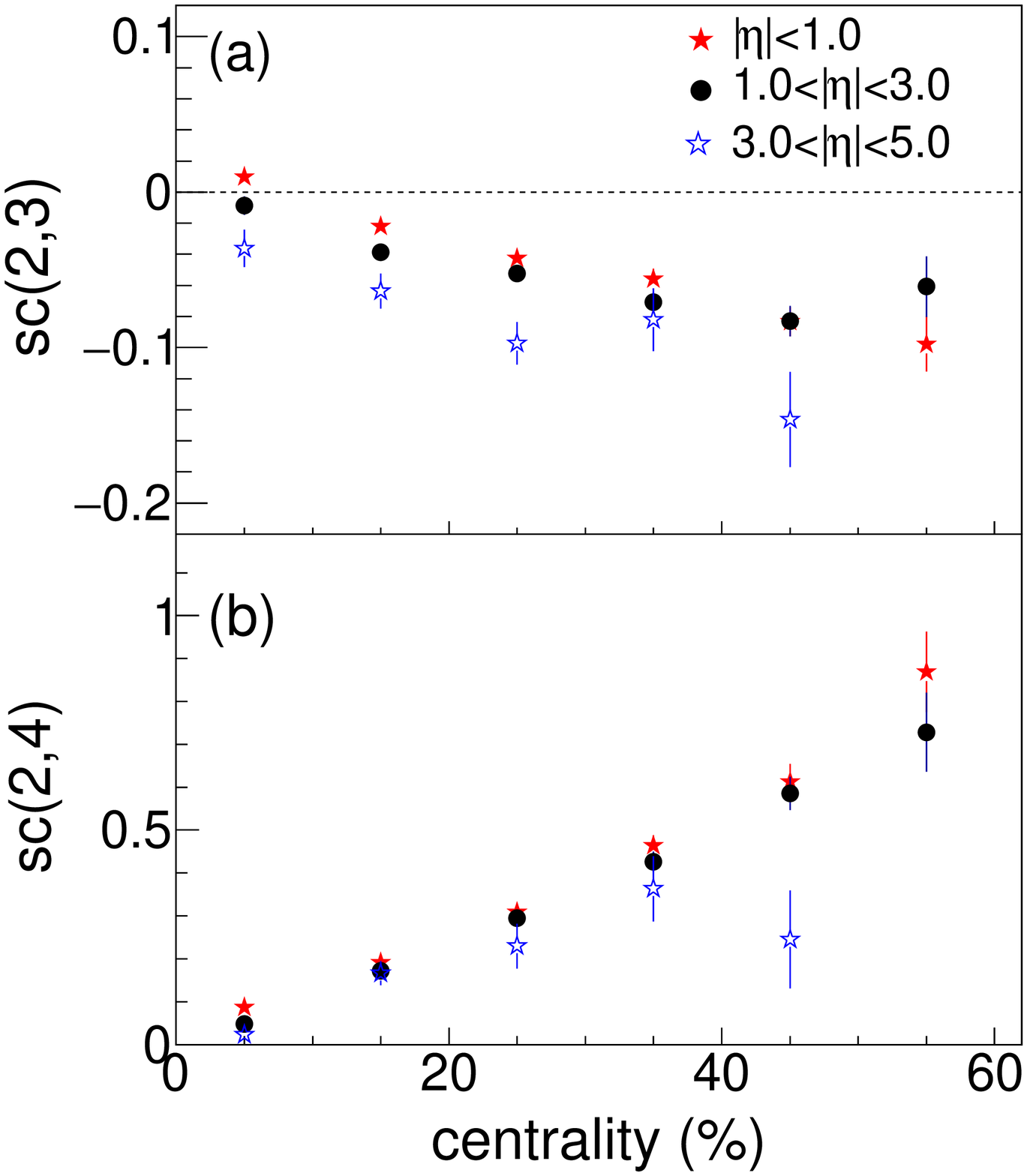}
\caption{(Color online) Centrality dependence of (a) $sc(2,3)$ 
and (b) $sc(2,4)$  in Au+Au collision at $\sqrt{s_{NN}}$ =200 GeV from
AMPT-SM model ($\sigma_{pp}$ = 3 mb) with $|\eta|<1.0$ (red) , $1 <|\eta|<3$ (black) and $3 <|\eta|<5$ (blue).}
\label{fig_scmn_rap}
\eef
\bef
\includegraphics[scale=0.35]{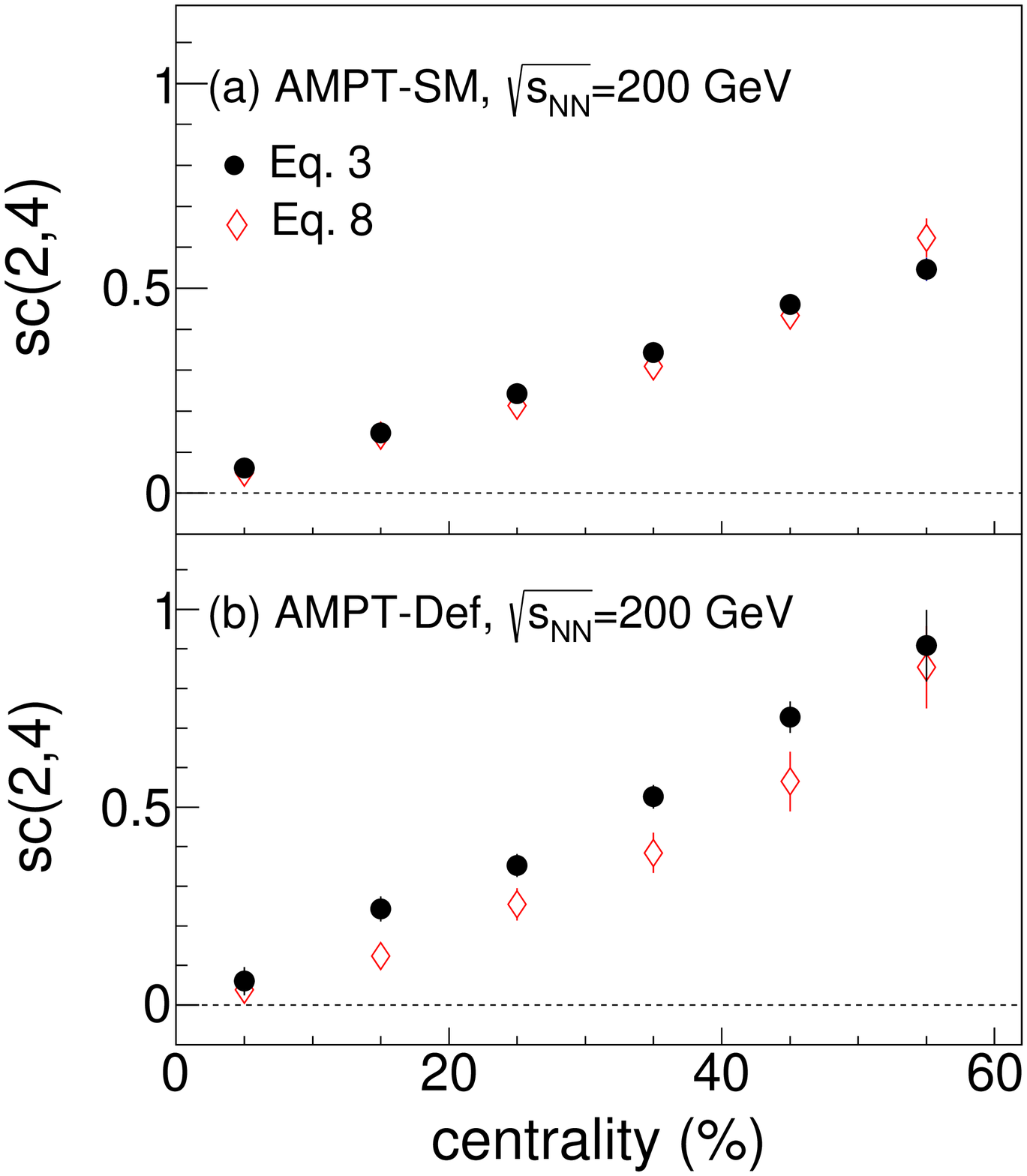}
\caption{(Color online) $sc(2,4)$ as a function of centrality from
  AMPT-SM ($\sigma_{pp}$ = 10 mb)  and AMPT-Def model in
Au+Au collision at $\sqrt{s_{NN}}$=200 GeV . Black (Red) symbol represents
calculation using Eq.~\ref{scmn_norm} (Eq.~\ref{scmn_phimn}) }
\label{fig_relation_sm_def}
\eef
\section{Relation between $sc(n,m)$ and Event plane correlation}
In ref~\cite{jaki}, authors derived a relation between $sc(n,m)$ and Event
plane correlation and validity of this relation is tested using
hydrodynamic calculation. Relation between $sc(2,4)$ and correlation
between 2nd and 4th order event plane ($\rm{cos}$ $\Phi_{24}$) is shown in Eq.~\ref{scmn_phimn},
\be
sc(2,4) = (\frac{\langle v_{2}^{6}\rangle }{\langle v_{2}^{4}\rangle
  \langle v_{2}^{2}\rangle }-1)\rm{cos}^{2} \Phi_{24},
\label{scmn_phimn}
\ee
where $\rm{cos}$ $\Phi_{24}\equiv \frac{Re \langle V_{4}(V_{2}^{*})^{2} \rangle}{\sqrt{\langle v_{2}^{2}\rangle\langle v_{2}^{4}\rangle}}$.
The magnitude of $\rm{cos}$ $\Phi_{24}$ is measured using Eq.~\ref{phimn}.
\be
\rm{cos} \Phi_{24} = \frac{\langle Q^{2}_{2A}Q^{*}_{4B}
  \rangle}{\sqrt{\langle Q_{4A}Q^{*}_{4B} \rangle} \sqrt{\langle
    Q^{2}_{2A} Q^{*2}_{2B}\rangle}   }.
\label{phimn}
\ee
Here $Q_{nA}$ and $Q_{nB}$  are the $n^{th}$ order flow vectors ($Q_n\equiv\sum_{k=1}^Me^{in\varphi_k}$) from two sub-event
separated in pseudorapidity. No gap between two sub-events has been
applied. The flow vectors $Q_{n}$  is calculated using charged particle  ($\pi$, $K$ and $p$) within 0.2
$<$ $p_{T}$ $<$ 2.0 GeV/c  and  0 $<$ $|\eta|$ $<$ 1.0. 
Eq.~\ref{scmn_phimn} relates 4-particle correlations ($sc(2,4)$) with
3-particle correlations (event-plane correlations) which is fully non
trivial. It has been observed that event-by-event hydrodynamics
satisfies Eq.~\ref{scmn_phimn} with a good approximation. Now, it is
worth to check whether  Eq.~\ref{scmn_phimn} is only valid in the hydro
framework or it can be true in transport model too. \\ 
The magnitude of $sc(2,4)$ has been measured as a function of centrality
using both Eq.~\ref{scmn_norm} and  Eq.~\ref{scmn_phimn} and  comparison
between them is shown in Fig.~\ref{fig_relation_sm_def}. Here the
magnitude  $sc(2,4)$ from Eq.~\ref{scmn_norm}  has been recalculated
by recalculating $v_{n}^{2}$ in the denominator using 2-particle method with
no pseudorapidity gap. This is done to make consistency between
Eq.~\ref{scmn_norm}  and Eq.~\ref{scmn_phimn}.
Fig.~\ref{fig_relation_sm_def}(a) shows the comparison between
Eq.~\ref{scmn_norm} and  Eq.~\ref{scmn_phimn} using AMPT-SM model in
Au+Au collision at $\sqrt{s_{NN}}$=200 GeV. The ratio
between red and black histograms is $\sim$ 20$\%$ in  AMPT-SM model. 
Fig.~\ref{fig_relation_sm_def}(b) shows comparison using AMPT-Def model. 
One finds that the deviation is larger in case of AMPT-Def (40$\%$)  compared to AMPT-SM (20$\%$).
I have also checked the validity of Eq.~\ref{scmn_phimn} by changing
hadronic cascade time and magnitude of $\eta_{s}/s$, however, the
conclusion remains the same. Therefore, Eq.~\ref{scmn_phimn}  which
relates 4-particle correlations with 3-particle
correlations is not valid in hadronic transport model, like AMPT-Def. It has been
shown in  ref~\cite{jaki} that the event-by-event hydro model satisfies
Eq.~\ref{scmn_phimn}. So, it is very important to check whether this
relation is also valid in real data or not.   \\

\section{Summary}
A set of predictions for the centrality dependence of the
normalized   symmetric cumulants in Au+Au collisions at $\sqrt{s_{NN}}$=200 GeV has been given using various
configuration of AMPT model.  
AMPT-Def, which is a hadronic model, shows a stronger (anti-) correlation between
$v_{2}$ and ($v_{3}) v_{4}$ compared to AMPT string melting
model. Effect of the shear viscosity on the magnitude of symmetric
cumulants ($sc(2,3)$ and $sc(2,4)$) is shown.  The magnitude of
$sc(2,4)$ and $sc(2,3)$ increases with increase in $\eta_s/s$. However,
$sc(2,4)$ is more sensitive to $\eta_s/s$ compared to $sc(2,3)$. 
The magnitude of $sc(2,3)$ is found to be sensitive with hadronic
rescattering, whereas  $sc(2,4)$ remains almost unaffected.  The
magnitude of  $sc(2,3)$ decreases with increase in hadronic
rescattering. Rapidity dependence of symmetric cumulants shows strong
(anti)-correlation between $v_{2}$ and ($v_{3}) v_{4}$ at mid-rapidity compared to forward/backward rapidity. 
A non-trivial relation between symmetric cumulant and event plane
correlation is tested in a transport-based AMPT model.  String melting
version of AMPT satisfy the relation, which connects  symmetric cumulant and event plane
correlation, with a better approximation compared to the AMPT-Def model( hadronic medium).\\
These new observables have not yet been measured at RHIC. These study
could be useful to understand upcoming measurements at
RHIC. 

\noindent{\bf Acknowledgments}\\
Financial assistance from the Department of Energy, USA is gratefully acknowledged. \\

\end{document}